\documentclass[final,5p,twocolumn,sort&compress]{elsarticle}

\usepackage{amssymb}

\usepackage[nodots]{numcompress}

\usepackage{multirow}
\usepackage{textcomp}
\usepackage{amsmath}
\usepackage[utf8]{inputenc}
\usepackage{graphicx}
\usepackage{overpic}
\usepackage{captdef}

\usepackage{color} 
\usepackage{epstopdf}

\journal{Computers \& Fluids}

\begin{document}

\begin{frontmatter}



\title{Complete PISO and SIMPLE solvers on Graphics Processing Units}


\author[PWr]{Tadeusz Tomczak}
\ead{tadeusz.tomczak@pwr.wroc.pl}

\author[PWr]{Katarzyna Zadarnowska}
\ead{katarzyna.zadarnowska@pwr.wroc.pl}

\author[UWr]{Zbigniew Koza}
\ead{zkoza@ift.uni.wroc.pl}

\author[UWr]{Maciej Matyka}
\ead{maq@ift.uni.wroc.pl}

\author[PWr2,corr]{{\L}ukasz Miros{\l}aw\corref{cor1}}
\ead{lukasz.miroslaw@vratis.com}

\address[PWr]{Institute of Computer Science, Control and Robotics, Wroc{\l}aw University of Technology, Wybrze{\.z}e Wyspia{\'n}skiego 27, 50-370 Wroc{\l}aw, Poland}

\address[UWr]{Faculty of Physics and Astronomy, University of Wroc{\l}aw, pl. M. Borna 9,
           50-204 Wroc{\l}aw, Poland}

\address[PWr2]{Institute of Informatics, Wroc{\l}aw University of Technology,
Wybrze{\.z}e Wyspia{\'n}skiego 27, 50-370 Wroc{\l}aw, Poland}

\address[corr]{Vratis Ltd, Muchoborska 18, PL54424  Wroc{\l}aw, Poland. Tel. +48 71 7073419, Fax. +48 71 7503104}

\cortext[cor1]{Corresponding author}

\begin{abstract}
We implemented the pressure-implicit with splitting of operators (PISO) and
semi-implicit method for pressure-linked equations (SIMPLE)
solvers of the Navier-Stokes equations
on Fermi-class graphics processing units (GPUs)
using the CUDA technology.
We also introduced a new format of sparse matrices optimized for performing elementary CFD operations, like
gradient or divergence discretization, on GPUs.
We verified the validity of the implementation on
several standard, steady and unsteady problems.
Computational efficiency of the GPU implementation was examined by comparing its double precision run times
with those of essentially the same algorithms implemented in OpenFOAM.
The results show that a GPU (Tesla C2070) can outperform a server-class 6-core, 12-thread CPU (Intel Xeon X5670) by a factor of 4.2.
\end{abstract}


\end{frontmatter}



\section{Introduction \label{sec:intro}}

As the silicon technology approaches subsequent physical barriers,
keeping the exponential growth rate of
computational power of computers
poses numerous scientific and technological challenges.
Today, most of performance improvements
comes from increased parallelism.
However, since a vast majority of the existing technologies for writing parallel applications were designed
for coarse-grained concurrency and rely on bulk-synchronous algorithms,
further progress requires new computer architectures, 
algorithms
and
programming models aimed at fine-grained on-chip parallelism
\cite{exascale}
A promising attempt in this direction
is the massively parallel architecture of modern graphics processor units (GPUs).
In just a few years GPUs evolved
into versatile programmable computing devices, whose
peak computational performance matches that of the most powerful supercomputers
of a decade ago.
For this reason GPUs are used as numerical accelerators on a vast variety of systems, from laptops to 
many of today's \cite{top500} and tomorrow's \cite{titan} petaflop supercomputers.

Many computational fluid dynamics (CFD) algorithms are inherently data-parallel,
and hence suitable for GPU acceleration.
There are, however, several obstacles
 to reach this goal.
First, existing industry-level CFD algorithms and data structures were mostly developed
for sequential or coarse-grained parallel architectures.
Second, the massively parallel  architecture of GPUs imposes
stiff conditions on the software to exploit the hardware efficiently.
Finally, redesigning existing applications and porting them to
GPUs using new programming paradigms requires a considerable time and effort.

In \cite{Malecha11} we accelerated
a standard finite volume CFD solver (OpenFOAM)
by replacing its linear system solvers and sparse matrix vector product
(SMVP) kernels with the corresponding GPU implementations.
A significant speedup was achieved only for steady state
problems, and we attributed marginal acceleration of transient problems to
frequent data format conversions and additional data transfers.
Here we tackle the problem of porting of a complete
finite volume solver
to the GPU.
We chose to implement  pressure-implicit with splitting of operators (PISO)
and semi-implicit method for pressure-linked equations (SIMPLE) solvers \cite{Ferziger-Peric}
and examine whether eliminating intermediate data transfers
through a narrow CPU-GPU bus
and adjusting the internal data format to the needs of the GPU
is sufficient to significantly accelerate the time to solution.
We also perform a more detailed analysis of the conditions necessary to
obtain high acceleration rates.


\section{Related work \label{sec:related-works}}

The literature devoted to porting CFD algorithms to GPUs is ample, but
a large part of the research
has focused on CFD solvers based on structured, regular grids.
While this approach facilitates coalescing of device memory accesses
and leads to a GPU-accelerated code that was reported as several \cite{cohen},
tens \cite{toelke-krafczyk} or even hundreds \cite{phillips} times faster than the corresponding CPU-only
solution, the usability of such CFD programs
is limited to
simple geometries or applications in computer graphics \cite{harris}.
For example, Cohen and Molemaker \cite{cohen} accelerated a solver for the Boussinesq approximation
of the Navier-Stokes equations on a regular three-dimensional (3D) grid, and found an 8-fold
speedup versus the corresponding multithreaded Fortran code running on an 8-core dual-socket Intel
Xeon processor. Another example is the work of T\"olke and Krafczyk \cite{toelke-krafczyk},
who found their GPU implementation of a
3D Lattice-Boltzmann method for flows through porous media
up to two orders of magnitude faster than a corresponding single-core CPU code.
While reports on two- or even three-digit speed gains should be interpreted cautiously \cite{Debunking},
they show a great potential of GPU accelerators.

Our GPU implementation works on \emph{unstructured} grids,
which is necessary to make it applicable for realistically complex geometries.
So far, optimized GPU implementations of CFD solvers based on unstructured grids have been relatively rare,
mostly because of stringent requirements for efficient utilization of the GPU.
For example, Kl\"ockner et~al.~\cite{klockner} implemented discontinuous Galerkin methods over unstructured grids
and found that the highest acceleration is achieved for higher-order cases due to their
larger arithmetic intensity which helps hide indirect addressing latencies.
Alternative optimization techniques were recently used by Corrigan et~al.~\cite{corrigan}
to show that a Tesla 10 series GPU can accelerate a 
finite-volume  solver for an inviscid, compressible flow over an unstructured grid
by almost $10$ times relative to an OpenMP-parallelized CPU code.

There are also several reports on accelerating existing CFD software with GPUs.
For example, Corrigan et~al.\ \cite{corrigan2,corrigan1}
used a Python script to semi-automatically translate OpenMP loops to GPU kernels
in a large-scale CFD code (nearly a million lines of Fortran 77 code), FEFLO.
Another strategy is to port complete solvers. This approach was used, for example, by
Elsen et~al.\ \cite{elsen}, who accelerated the Navier Stokes Stanford University Solver (NSSUS)
and found the speedup of over $20$ times for complex geometries containing  up to $1.5$ million grid points.
Porting of the MBFLO2 multi-block turbulent flow solver
was described by Phillips et~al.\ \cite{davis}, who found a 9-fold speedup over the original CPU implementation.
Both FEFLO and MBFLO2 libraries are, however, based on structured grids.

While there are several libraries aimed at GPU-acceleration of existing unstructured-grid CFD libraries, e.g.\
Cufflink (http://cufflink-library.googlecode.com) and  ofgpu (http://www.symscape.com/gpu-0-2-openfoam),
their design adheres to the ``partial acceleration'' rather than ``full port'' strategy.
In particular, they typically use GPUs only to accelerate some well-defined, time-consuming and data-parallel tasks, like
solving large sparse linear systems. This approach introduces some artificial overhead related to
frequent CPU $\to$ GPU $\to$ CPU data transfers as well as data format conversions.
Therefore our aim is to go a step further and develop a GPU-only implementation of a CFD solver
that would have selected features of industrial solvers:
support for arbitrary meshes (orthogonal or nonorthogonal) and time-dependent boundary conditions.


\section{Implementation\label{sec:Implementation}}

PISO and SIMPLE are standard CFD solvers for incompressible flows and
we followed \cite{JasakPHD,OpenFOAM} in their
implementation. 
We used the finite volume method (FVM) to
discretize Navier-Stokes equations
which are then solved iteratively until convergence.
This iterative procedure consists of a
sequence of well-defined steps. For example, PISO starts with the
solution of the momentum equation followed
by a series of solutions of the pressure equation and explicit
velocity corrections.
The SIMPLE model works on similar principles, but is optimized for
steady-state flows.
Thus, at such a coarse-grain level of description, both algorithms are
essentially sequential
and cannot be parallelized. Fortunately, individual steps of these
algorithms can be parallelized using GPU.

To port the solvers to the GPU architecture, we used
CUDA, the computer architecture and software development tools for modern
Nvidia accelerators \cite{Garland10,Farber2011}. One of the most
distinguishing features
of the CUDA programming model is the hierarchical organization of the memory.
Thus, one of major contributions of our paper is reorganization of solver
data structures
to enhance the efficiency of internal data transfers in a GPU device.
In particular, we focused on enhancing efficiency of  the solution of
large sparse linear systems,
the most time-consuming operation  in the PISO and SIMPLE solvers.
This nontrivial problem is the subject of intensive research
\cite{nvidia:precond-iter,koko12,Geveler2012}, as many  advanced techniques,
like LU-based preconditioning, contain large serial components. Here
we focus on conjugate gradient (CG) and
biconjugate gradient stabilized (BiCGStab) iterative solvers with
Jacobi preconditioning \cite{Barrett94},
two methods known to be amenable to effective fine-grained parallelization.


\subsection{Data format}

Since GPUs not only execute programs in parallel, but also access their memory in parallel,
choosing right data structures is of highest importance.
The data processed in FVM-based CFD solvers come from discretization of space, time and
flow equations. Space is divided into a mesh of $N$ cells. Cells are polyhedrons with flat faces, and each face
belongs to exactly two polyhedrons or is a boundary face. Pressure and velocity
are defined at centroids of the cells. Since partial differential equations are local in space and time,
their discretization leads to nonlinear algebraic equations relating the velocity and pressure at each
polyhedron with their values at adjacent cells only. After linearization, these equations reduce to a linear system
$$
  \hat{A}\vec{x} = \vec{b},
$$
where $\vec{x}$ and $\vec{b}$ are vectors of length $N$ and $\hat{A}$ is a sparse matrix such that
$A_{ij} \neq 0$ if and only if cells $i$ and $j$ have a common face or $i = j$.
The value of $A_{ij}$ can depend on the current and previous values of the pressure and velocity at $i$, $j$
as well as on some face-specific parameters, e.g.\
area of the face. 
Matrix $\hat{A}$ must be assembled many times and then used in a linear solver. During these operations
$\hat{A}$ is accessed in rows or columns as if in sparse matrix-vector and sparse transposed matrix-vector products (STMVP).
Although the highest priority must be granted to the optimization of SMVP,
$\hat{A}$ must be at the same time stored in a way enabling a
reasonably efficient implementation of STMVP.

\begin{figure}
 \centering
  \includegraphics[width=0.9\columnwidth]{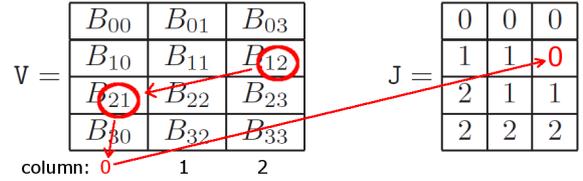}
    \caption{A schematic of evaluation of \texttt{J[1][2]} for the examplary matrix described in the text.
      First it is checked \texttt{V[1][2]} hold $B_{12}$.
      Then the indices of $B_{12}$ are transposed to get $B_{21}$ and since the latter is stored in \texttt{V} in column 0,
      \texttt{J[1][2]} $ = 0$.}
    \label{fig:J}
\end{figure}

Several  formats designed for efficient implementation of the SMVP kernel on modern GPUs were investigated by
Bell and Garland \cite{Bell09} and the data format implemented in our SIMPLE and PISO solvers
is similar to their hybrid format. In the original hybrid format $\hat{A}$ is split into two disjoint parts:
$\hat{A} = \hat{B} + \hat{C}$, where $\hat{B}$ is stored in the ELL format, whereas
$\hat{C}$ is stored in the COO format.
Consider the following example:
$$
  \hat{B} = \left(
  \begin{array}{cccc}
    B_{00} & B_{01} &   0    & B_{03}  \\
    B_{10} & B_{11} & B_{12} & 0       \\
    0      & B_{21} & B_{22} & B_{23}  \\
    B_{30} &   0    & B_{32} & B_{33}
  \end{array}
  \right)
$$
In ELL, this matrix would be represented in two 2D arrays: \texttt{V} and {I} \cite{Barrett94},
$$
 \mathtt{V} =
  \begin{array}{|c|c|c|}
  \hline
    B_{00} & B_{01} & B_{03} \\
  \hline
    B_{10} & B_{11} & B_{12} \\
  \hline
    B_{21} & B_{22} & B_{23} \\
  \hline
    B_{30} & B_{32} & B_{33}\\
  \hline
  \end{array}\;,
        \qquad
 \mathtt{I} =
  \begin{array}{|c|c|c|}
  \hline
    0 & 1 & 3 \\
  \hline
    0 & 1 & 2 \\
  \hline
    1 & 2 & 3 \\
  \hline
    0 & 2 & 3\\
  \hline
  \end{array} \;.
$$
In general, dimensions of both of these arrays are $N\times K$, where $N$ is the number of cells,
$K$ is a small integer, and the entries in \texttt{I}
are integers between 0 and $N-1$.
While this format gives excellent memory bandwidth when the matrix is accessed by rows,
it is completely inadequate for accessing it by columns.
We can, however, take advantage of the fact that $\hat{A}$ is structurally symmetric ($A_{ij}\neq0$ iff
$A_{ji}\neq 0$)
and extend the ELL format with an additional  $N\times K$ integer array \texttt{J}
defined indirectly as follows.
For the sake of clarity assume that $\hat{B}$ is also structurally symmetric.
By definition of ELL, for any $0 \le i < N$ and $0 \le k < K$, $ V_{ik} = B_{ij}$ with
$j$ = $I_{ik}$.
The corresponding entry in the transposed matrix, $B_{ji}$, is stored in \texttt{V} in row $j$ and some column $k'$ ($0 \le k' < K$).
The value of $J_{ik}$ is defined as  $k'$.
Evaluation of one entry of \texttt{J} for the exemplary matrix $\hat{B}$ is illustrated in Fig.~ \ref{fig:J}, 
and its full form reads
$$
 \mathtt{J} =
  \begin{array}{|c|c|c|}
  \hline
    0 & 0 & 0 \\
  \hline
    1 & 1 & 0 \\
  \hline
    2 & 1 & 1 \\
  \hline
    2 & 2 & 2\\
  \hline
  \end{array} \;.
$$

To access $\hat{B}$ by rows one can simply use the ELL format, i.e., arrays \texttt{V} and \texttt{I}.
To access a column $k$ of $\hat{B}$, one first reads the entries in row $k$ of array \texttt{I}, which in this case identify
the row numbers of nonzero entries in this column.  These row numbers are then used as row indices into \texttt{V},
with column indices read from \texttt{J}.

If $\hat{B}$ is not structurally symmetric, i.e. if $B_{ij} \neq 0$ and $B_{ji} = 0$ for some $i,j$,
then the value of $A_{ji}$ is stored in $C$ $(C_{ji} = A_{ji})$ and $A_{ij}$ is stored at $V_{ik}$ with some $0\le k < K$.
This can be indicated in the computer representation by assigning a negative value to $J_{ik}$.
In a similar way  the computer representation of $C$ stores the information that $A_{ij}$ is to be found in $\hat{B}$.

Note that while arrays \texttt{I} and  \texttt{J} facilitate column accesses to a matrix,
these accesses rely on indirect addressing, which results in
uncoalesced, inefficient memory transfers. The only way to improve this would be to store a transposed matrix
besides the original one,  but this would require not only extra storage, but also
a costly matrix transposition to be repeated every time
a new matrix is assembled.
In contrast, arrays \texttt{I} and \texttt{J} can be initialized only once in the program and
be shared by all data related to the faces.
Note also that arrays \texttt{I} and \texttt{J} could be combined into a single integer array, \texttt{Q},
to reduce memory requirements of the program. For example, $I_{ij}$ and $J_{ij}$ can be encoded in $Q_{ij}$
as $N\times$\texttt{I}$_{ij} + $
\texttt{J}$_{ij}$ or $K\times$\texttt{J}$_{ij} + $
\texttt{I}$_{ij}$, with trivial decoding of $I_{ij}$ and $J_{ij}$ through integer division and remainder operations.

While in the original hybrid format $\hat{C}$  is represented in the COO format,
we used compressed row sparse (CRS) instead to further reduce the memory usage.
However, this particular choice does not seem to influence the overall performance significantly.
Most of the polyhedrons forming the mesh are of the same simple shape, e.g., they are tetrahedra or parallelepipeds.
This means that most cells have the same number of neighbours, typically 4 or 6. This, in turn, implies that
most of the entries in $\hat{A}$ can fit into $\hat{B}$, with $\hat{C}$ containing only a small fraction
of nonzero entries in $\hat{A}$. In particular, $\hat{C}$ usually vanishes altogether for structured meshes.


\section{Results \label{sec:results}}


\subsection{Test cases\label{sub:test-cases}}
\begin{figure*}
\centering
 \includegraphics[width=\textwidth]{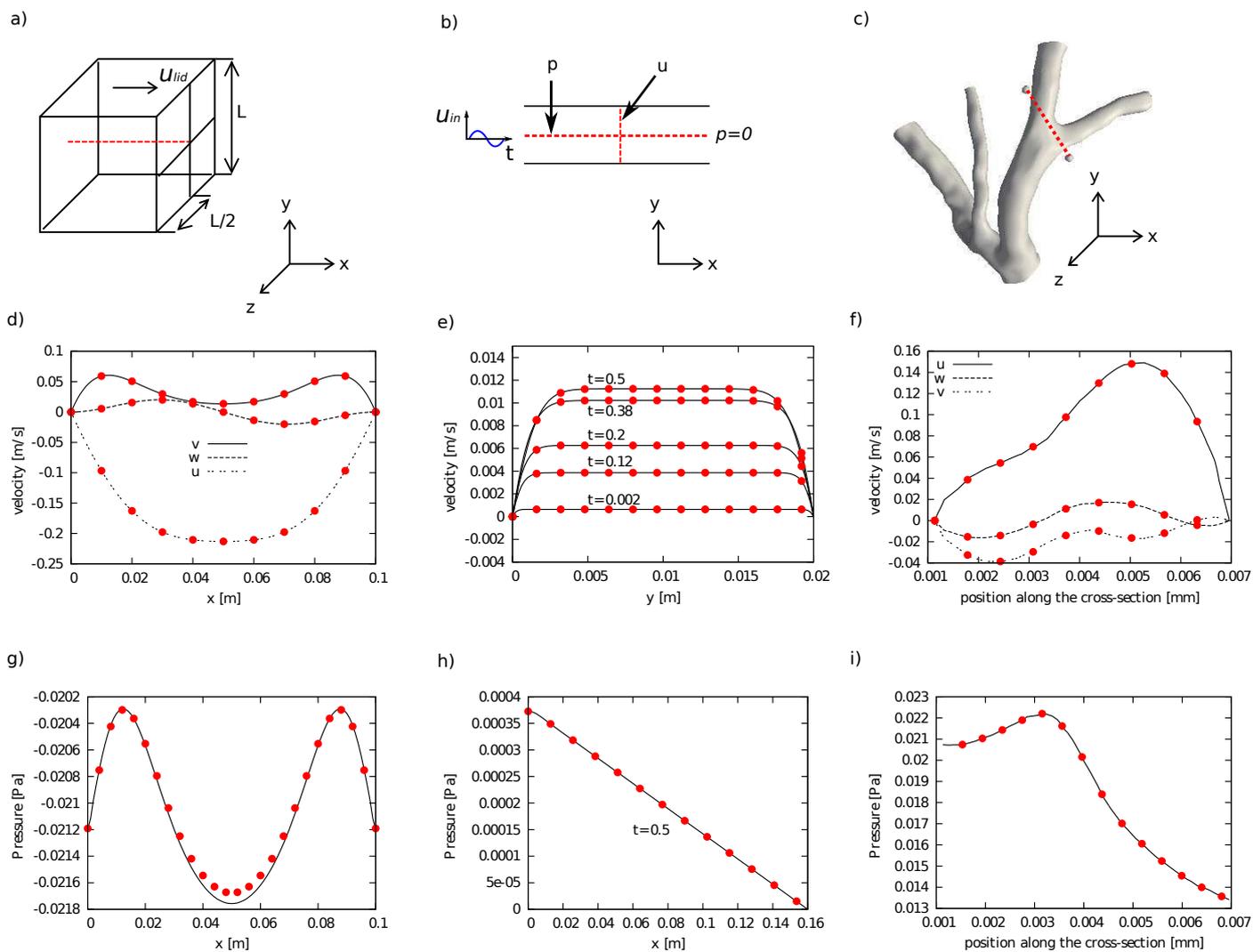}
 \caption{
  (Color online) Validation cases: a) lid-driven cavity ($10^6$ cells), b) transient Poiseuille flow,
  c) coronary artery. The dotted (online: red) line is the cross-section along which
  the results are evaluated. Results: d,f) velocity components $u$, $v$, $w$ (parallel to the $x$, $y$ and $z$ axis, respectively)
  along the cross-section; e) velocity $u$ at several times (in seconds);
  g,h,i) pressure along the cross-section.
  The results obtained on the CPU and GPU are shown as symbols
  (\textcolor{red}{$\bullet$}) and solid lines, respectively.
  Only part of the available CPU data are shown for clarity.
 \label{fig:res}}
\end{figure*}

To validate the code and evaluate its performance we solved three CFD
problems: a steady flow in a 3D lid-driven cavity
\cite{Shankar00},  the transient Poiseuille flow \cite{Womersley55} in
two dimensions,
and the steady flow through the human left coronary artery (LCA) (see
Fig.~\ref{fig:res} a--c).
The cavity was a cube of side length $\textrm{L}=0.1$~m.
A constant velocity $u_\mathrm{lid} =
1$~m/s was imposed at its top face and the kinematic viscosity $\nu=0.01$~m$^2$/s
was assumed (Reynolds number ${\mathrm{Re}=10}$).
To evaluate how the solver performance depends on the problem size,
we varied the regular mesh resolution from
$10^3$ to $223^3$ cells (see Tab.~\ref{tab_test_cases}).

\begin{table}
\caption{Basic parameters of the meshes}
\label{tab_test_cases}
\centering
\begin{tabular}{ l  r  r  c }
\hline
Case name          & Cells         & Faces         & Structured? \\
\hline
cav$10^3$   & $     1~000$  & $     3~300$  & yes \\ 
cav$47^3$   & $   103~823$  & $   318~096$  & yes \\ 
cav$100^3$  & $ 1~000~000$  & $ 3~030~000$  & yes \\ 
cav$181^3$  & $ 5~929~741$  & $17~887~506$  & yes \\ 
cav$223^3$  & $11~089~567$  & $33~417~888$  & yes \\ 
tp$1$M     & $   960~000$  & $ 3~843~500$  & yes \\ 
lca$0.5$M   & $   458~861$  & $ 1~071~019$  &  no \\
\hline
\end{tabular}
\end{table}

In the transient Poiseuelle flow (\texttt{tp1M} in Tab.~\ref{tab_test_cases})
we defined a 2D pipe of length
$0.16$~m and height $0.02$~m and assumed
$\nu=3.3\times10^{-6}$~m$^2$/s.
The mesh resolution was $3200\times 300$ cells.
We assumed a time-dependent inlet velocity:
\begin{equation*}
u_\mathrm{in}(t) = u_0 \sin(2\pi f t)
\end{equation*}
with $u_0 = 0.01$~m/s and $f= 0.5$~s$^{-1}$, and
the zero-pressure boundary condition at the outlet.
In this case we ran $5000$ PISO steps with $\delta
t=0.1$~ms.

Geometry of the artery simulated in the \texttt{lca0.5M} case (Tab.~\ref{tab_test_cases}) is visualised
in Fig.~\ref{fig:res}c. We used
$\nu=3\cdot10^{-6}$~m$^2$/s and assumed a constant mass flow at the inlet ($Q = 9.975\times
10^{-4}$ kg/s). We imposed the zero-pressure
boundary conditions in all the outlets and no-slip boundary conditions
at artery walls.
A non-uniform and non-orthogonal mesh built of $458~861$ cells was
used. To solve this case,  we ran $350$ iterations of SIMPLE and $2000$ iterations of PISO
solvers, the latter with $\delta t=0.1$~ms.

\subsection{Validation}

To validate our GPU implementations we compared its
solutions for all  scenarios described in  Sec.~\ref{sub:test-cases}
to the results obtained with the OpenFOAM toolkit  \cite{OpenFOAM} running on the CPU.
OpenFOAM was run with the same internal algorithms and control
parameters as the GPU code except for  the sparse unsymmetric linear solver:
in the GPU code we used BiCGSTAB, whereas OpenFOAM uses BiCG.
All computations were done in double precision.

Results are depicted in Fig.~\ref{fig:res}. Panels (a), (b), and (c) show the geometry of the test cases:
cavity, Poiseuille and LCA, respectively.
Panels (d) and (f) show the Cartesian components of the velocity, panel (e) presents the velocity
component parallel to pipe walls at several times, and  panels (g), (h), and (i)
show the pressures.
Data for the cavity and LCA were computed using the SIMPLE solver, whereas for
Poiseuille problem PISO was used.

In general, one can see very good matching of the GPU and CPU
results.
Only a slight disagreement in the minimum value of
the pressure in a cavity flow was found (see Fig.~\ref{fig:res}g).
This discrepancy is marginal (results differ on the fourth significant digit),
and we attribute it to the fact that different solvers for nonsymmetric
sparse systems were used.


\subsection{Performance evaluation}

Performance tests of the GPU code
were done on the Tesla C2070 graphics accelerator attached to
a PC running 64-bit Ubuntu 10.04 
LTS, graphics driver v.~290.10, Nvidia CUDA 4.1 and gcc 4.4.3.
The reference CPU tests were performed using OpenFOAM v.~1.7
on the Intel Xeon X5670 processor, which is a 6-core chip with hyperthreading.
For this reason  we compared a single GPU accelerator with a single CPU processor running 12 threads.
Moreover, since our GPU code relies on a very simple Jacobi (diagonal) preconditioning,
we also repeated all CPU tests using the geometric-algebraic multi-grid solver (GAMG),
which is considered to be among the fastest solvers of the pressure equation available in OpenFOAM.


\subsubsection{Time to solution}

We define the solver performance as the total time to solution, including the time necessary
to read the test case definition from a disk.
To find it, we ran GPU and CPU (OpenFOAM) solvers
until residuals in the pressure and velocity solvers dropped below a given threshold, typically $10^{-5}$
for smaller and $10^{-4}$ for larger cases.
Depending on the problem size, this time varied from 1 second to almost 31 hours.
Similarly, we define the effective GPU acceleration as the ratio of the GPU solver performance to its CPU counterpart.

\begin{figure}
\centering
\includegraphics[width=0.9\columnwidth]{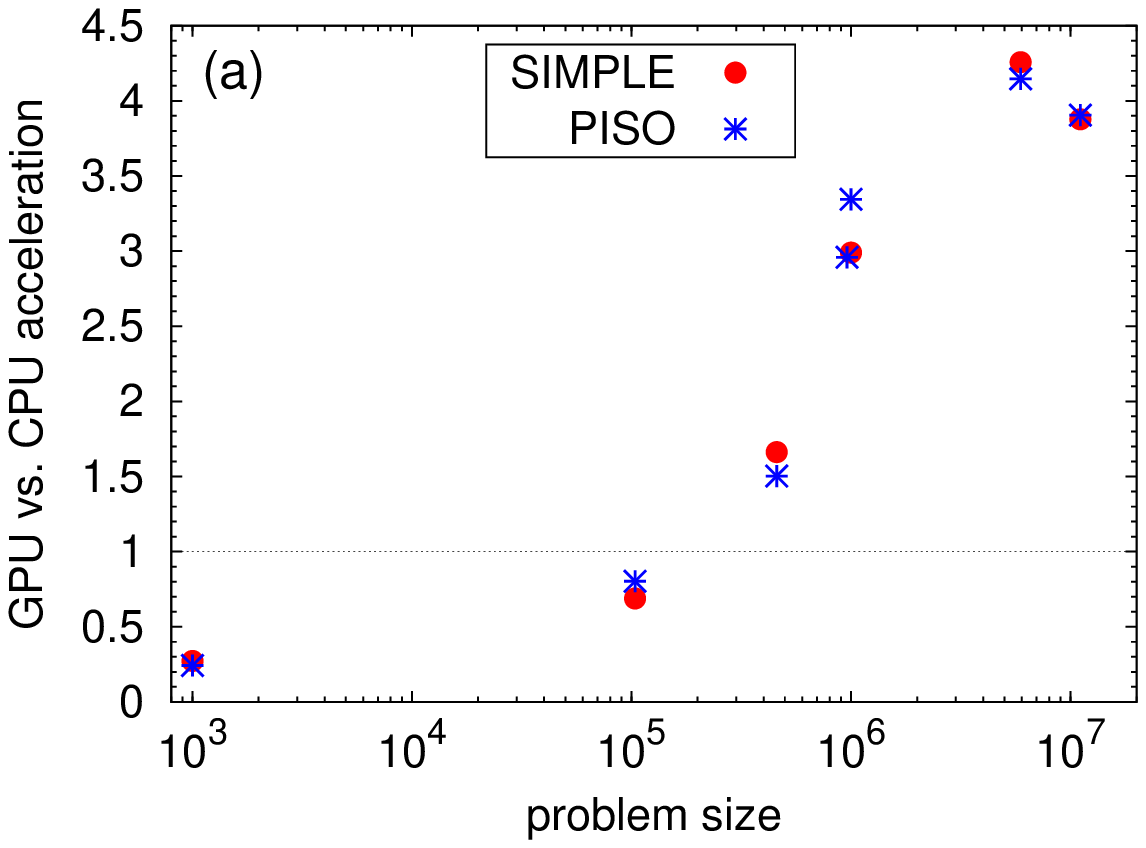}\\
\includegraphics[width=0.9\columnwidth]{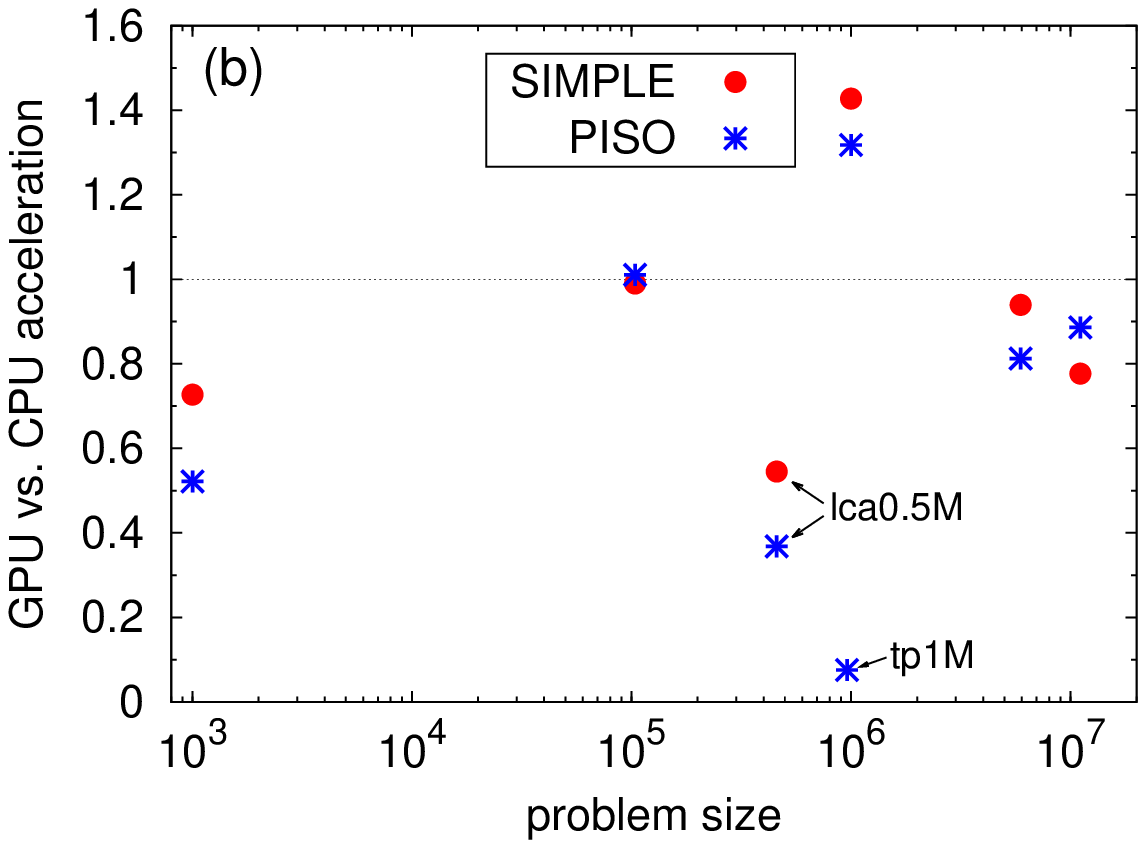}\\
\caption{Acceleration of the GPU relative to the CPU (OpenFOAM) implementations of SIMPLE and PISO solvers.
In all cases the linear solvers used by the GPU were Jacobi-preconditioned
BiCGStab and CG. The CPU implementation used BiCG and CG with Jacobi preconditioner  (a)
or GAMG with the FDIC preconditioner (b).
}
\label{fig:accel-comparison}
\end{figure}

Figure \ref{fig:accel-comparison}a shows the effective acceleration
when the GPU and CPU solvers use essentially  the same algorithms and control parameters,
the only exception being the BiCG solver rather than BiCGStab used by OpenFOAM to solve the velocity equations.
As expected, in this case the relative performance of the GPU and CPU implementations depends strongly on the problem size.
The GPU is slower than the CPU if a mesh has less than approximately $N = 10^5$ cells, and is
significantly faster, up to about 4.2 times, if $N$ is of order of a million or more.
This reflects the fact that the GPU is a massively parallel device that needs tens of thousands of threads for efficient
program execution.
One can also see that the relative GPU-CPU performance is similar for the PISO and SIMPLE solvers.

Comparison of the GPU implementations with OpenFOAM exploiting  the GAMG pressure solver
is depicted in Fig.~\ref{fig:accel-comparison}b. In this case the GPU can deliver
a small performance gain (about 30\%--40\%) only for the \texttt{cav$100^3$} case.
The major factor affecting the relative GPU-CPU performance
is now the structure of the problem rather than just its size:
GAMG turns out much more efficient for \texttt{lca0.5M} and \texttt{tp1M} cases
than for the driven cavity. This is directly related to the fact that for the former problems
a Jacobi-preconditioned CG solver used in the GPU implementation needs a large number of iterations to converge, see
Tab.~\ref{tab_CG_nr_iter}. In the extreme case of the transient Poiseuille flow, \texttt{tp1M}, the CG solver
requires almost $700$ times more iterations than  GAMG.
Even though GAMG iterations are far more complex, a GAMG-based PISO solver running on the CPU
turns out  $13$ times faster than its  CG-based counterpart running on the GPU.

\begin{center}
\begin{table}
\caption{
	Cumulative number of the pressure solver iterations: Jacobi-preconditioned CG (GPU and CPU) and GAMG (CPU).
}
\label{tab_CG_nr_iter}
\centering
\begin{tabular}{ l  r  r  r  r  r  r }
	\hline
Case                    & GPU-CG             & CPU-CG            & CPU-GAMG  \\
 \hline
    \multicolumn{4}{c}{SIMPLE}                    \\
	cav$10^3$           & $    2~468$        & $    2~474$       &  $   269$         \\
	cav$47^3$           & $  122~216$        & $  103~321$       &  $ 3~699$         \\
	cav$100^3$          & $  635~086$        & $  629~182$       &  $14~721$         \\
	cav$181^3$          & $1~424~531$        & $1~618~634$       &  $29~191$         \\
	cav$223^3$          & $1~733~366$        & $1~766~484$       &  $27~220$         \\
	lca$0.5$M           & $  157~176$        & $  159~959$       &  $ 1~704$         \\
\hline
    \multicolumn{4}{c}{PISO}                    \\
	cav$10^3$          &   $  104~972$    & $   110~614$    & $ 10~507$     \\
	cav$47^3$          &   $  575~153$    & $   568~388$    & $ 16~549$     \\
	cav$100^3$         &   $1~255~664$    & $ 1~352~614$    & $ 25~257$     \\
	cav$181^3$         &   $2~726~289$    & $ 3~010~947$    & $ 43~760$     \\
	cav$223^3$         & $  3~705~806$    & $ 3~716~480$    & $100~279$     \\
	lca$0.5$M          & $  1~770~672$    & $ 1~830~437$    & $  9~875$     \\
	tp1M               & $ 24~117~157$    & $27~042~577$    & $ 34~589$     \\
\hline \end{tabular}
\end{table}
\end{center}

The values in the second and third column of Tab.~\ref{tab_CG_nr_iter}
differ by up to about 10\% even though they refer to essentially the same quantity.
This reflects the fact that the GPU and CPU implementations are not equivalent, as they use different
linear solvers for velocity equations. Moreover, the order in which floating point operations are executed
in both architectures is different, which leads to architecture-specific accumulation of numerical errors
and divergent program execution.


\subsubsection{Profiling}

Profiling large-scale GPU code can be  quite problematic.
On the one hand, CPU-oriented profilers, like oprofile, usually find it difficult to distinguish calls to different CUDA functions.
On the other hand, CUDA-specific profilers, like Nvidia Visual Profiler, are designed
to provide fine-grained information on kernel
execution (such as branch divergence or non-coalesced device memory accesses), and can incur prohibitive runtime overhead.
Therefore we used a different strategy---since each GPU kernel must be launched from the CPU host,
we modified the CPU source code by inserting calls
to \texttt{gettimeofday} system timer.

\begin{figure}
\centering
\includegraphics{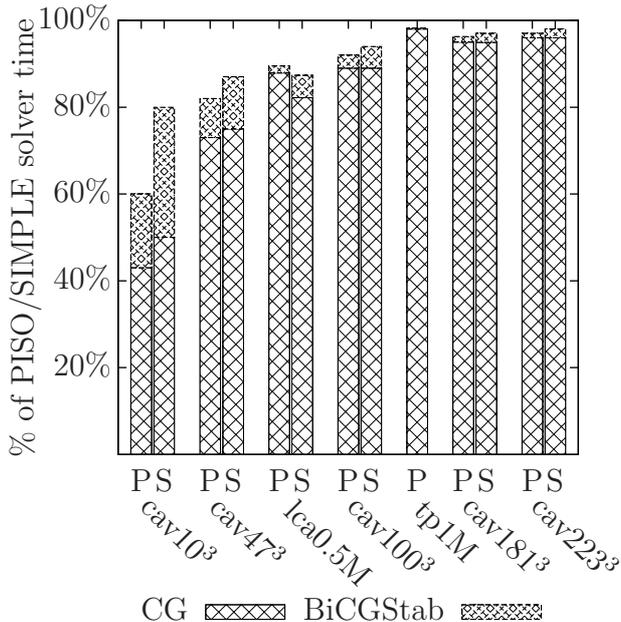}
\caption{
	Percentage of the time spent in CG and BiCGStab linear solvers during a full PISO (P) and SIMPLE (S) solver execution.
}
\label{fig_CG_in_total}
\end{figure}

Figure \ref{fig_CG_in_total} shows the fraction of the total computation time the GPU-accelerated PISO and SIMPLE solvers spent
in CG and BiCGStab linear solvers.
As expected, the Jacobi-preconditioned CG solver turned out the most time-consuming part of our CFD solvers,
a single factor limiting the overall performance in all but the smallest cases.
Other major components of our implementation, including the Jacobi-preconditioned BiCGStab solver of
sparse nonsymmetric linear systems, are negligible from the performance point of view.
This is no surprise, as our implementation of the CG solver uses the simplest nontrivial preconditioner,
characterized by a poor convergence rate. This, in turn, can lead
to huge numbers of internal CG iterations listed in Tab.~\ref{tab_CG_nr_iter}.

\begin{figure}
\centering
\includegraphics{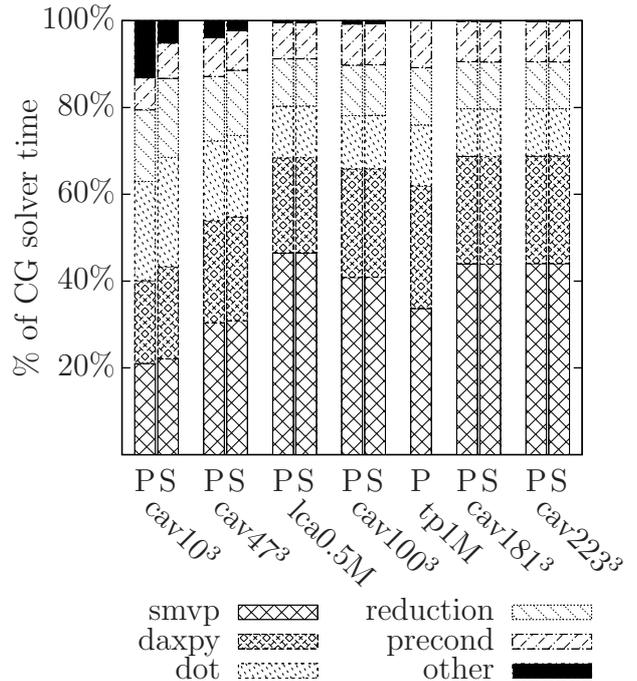}
\caption{
	Percentage of the time spent in the main components  of the Jacobi-preconditioned CG solver
    PISO (P) and SIMPLE (S).
}
\label{fig_CG_prof}
\end{figure}

As the data in Fig.~\ref{fig:accel-comparison} show that in some cases our GPU implementation
is competitive to an algorithmically different, highly optimized CPU implementation,
one might ask whether it is possible to improve the relative GPU-CPU performance simply by
improving the GPU implementation of the Jacobi-preconditioned CG solver.
To answer this question, we profiled the CG solver and show the results in Fig.~\ref{fig_CG_prof}.
Six groups of elementary numerical operations were investigated:
SMVP (sparse matrix dense vector multiplication),
daxpy (scalar times vector plus vector, $\vec{y} = \alpha \cdot \vec{x} + \vec{y}$),
dot product, reduction,
diagonal preconditioner, and all other operations (e.g.\ memory allocation and management).
Since all five explicitly listed operations take at least 10\% of the total CG solver time,
significant improvement of the solver performance would require
significant improvements in practically all of the solver components.
However, as many of them (SMVP, daxpy, dot product) are memory-bound and are already highly optimized,
no further significant improvement seems possible without additional hardware support.


\subsubsection{Matrix assembly\label{subsub:matrix-assembly}}
While the time to solution for the cases considered in this study turned out completely dominated
by the CG solver, this need not be so in other cases, other implementations, or other CFD solvers.
Therefore we also profiled elementary steps of the linear matrix assembly stage.
We considered three such operations: Laplacian, divergence, and gradient and compared their
execution times with that of the SMVP, as all these operations are algorithmically similar.
The results are collected in Tab.~\ref{tab_oper_time}.
\begin{table}
\caption{
   Computation time of elementary matrix assembly operations in SIMPLE and PISO solvers.
   The data for the divergence, gradient and Laplacian are normalized relative to the SMVP time.
   The last column shows orthogonality of the mesh.
}
\label{tab_oper_time}
\centering
\begin{tabular}{ l  r  r  r  r  c}
\hline
 \multirow{2}{*}{Case}         & \multicolumn{1}{c}{SMVP time } & \multicolumn{3}{c}{Normalized time} & \multirow{2}{*}{Orth.} \\
  \cline{3-5}
              & \multicolumn{1}{c }{[{\textmu}s]} &  div. & grad. &  Lap. &\\
              \hline
cav$10^3$   &      $41$  &  $3.6$  &  $16.8$ &  $ 2.3$   & yes\\
cav$47^3$   &     $139$  &  $4.1$  &  $18.3$ &  $ 2.5$   & yes\\
cav$100^3$  &   $1~103$  &  $2.9$  &  $10.7$ &  $ 2.4$   & yes\\
cav$181^3$  &   $6~763$  &  $2.5$  &  $ 9.1$ &  $ 2.3$   & yes\\
cav$223^3$  &  $12~620$  &  $2.5$  &  $ 9.0$ &  $ 2.3$   & yes\\
lca$0.5$M   &     $708$  &  $2.6$  &  $12.1$ &  $58.7$   & no\\
tp1M        &     $775$  &  $2.8$  &  $10.4$ &  $ 2.3$   & yes\\
\hline
\end{tabular}
\end{table}
On can see that for orthogonal meshes the matrix assembly time is dominated by the gradient.
This is a GPU-specific phenomenon, a consequence of the fact that in this operator the faces are
accessed in a different order than in SMVP, which leads to uncoalesced memory accesses and significant
overall performance degradation. For nonorthogonal meshes the Laplacian requires additional, costly corrections
and dominates the matrix assembly time.


\subsubsection{Memory usage}

The largest problem we were able to run on a 6~GB device, \texttt{cav223$^3$},
occupied 5.9~GB for the PISO and 5.7~GB for the SIMPLE solver, respectively.
In each test case the device memory footprint was similar for PISO and SIMPLE and,
for problems of size $\gtrsim 10^5$, equaled to  $\approx 540$ bytes per cell.


\section{Discussion \label{sec:discussion} }

Our results show that a GPU can outperform a six-core server-class CPU running algorithmically
equivalent implementations of popular CFD solvers,
SIMPLE or PISO, by a factor slightly exceeding 4.
This value is consistent with the  ratio 4.5 of the theoretical memory bandwidths
of the two processors used in our test, 144 GB/s (GPU) and 32 GB/s (CPU).

We show, however, that despite its huge computational power,
the GPU is still inferior to the CPU if the latter uses the most efficient algorithms.
This is because the GPU is not flexible enough to allow both direct and
efficient implementation of many procedures optimized for CPUs.
For example, some of the most efficient preconditioners for the CG sparse linear solver
are based on incomplete LU decomposition, which, unfortunately,  has a large serial component.
This issue has attracted a lot of interest. Recently, Naumov showed a $2\times$ speedup over a quad-core CPU in the incomplete-LU
preconditioned iterative methods \cite{nvidia:precond-iter},
Helfenstein and Koko \cite{koko12} reported a 10-fold acceleration over a single-threaded CPU code, and
Nvidia included ILU0-class preconditioners in its forthcoming cusparse 5.0 GPU library \cite{cusparse5}.
Even better results for the pressure solver performance can often be achieved with multigrid methods
and the effort to produce efficient GPU implementation of such solvers is also intensive,
see for example Geveler et~al.\ \cite{Geveler2012} and references therein.
In particular, Geveler et~al.\ reported a 5$\times$ average speedup over a multithreaded
CPU code.
These achievements are mostly concentrated on accelerating a particular, rather narrow aspect of a CFD solver.
By replacing  in our code the Jacobi-peconditioned CG with one of the above-mentioned solutions, it should be
possible to produce a fully functional CFD solver that would at least
match the speed of standard CFD software running on CPUs, which is our plan for the future.

If a faster pressure solver is used in future CFD software on GPUs, the role of the remaining solver components will
increase dramatically.
Moreover, since the gap between the computational power of processors and the speed of the bus connecting GPU with CPU
is expected to be still widening, porting of complete software rather than accelerating only its parts will
become more and more desirable.

\section{Conclusions \label{sec:conclusions} }

We implemented PISO and SIMPLE solvers on GPUs and investigated their main properties.
The implementations are fully functional, execute completely on the GPU using double precision
and support time-dependent boundary conditions and arbitrary meshes.
To facilitate a complete GPU port, we proposed a generic data format for internal data storage which
helps implement elementary CFD solver operations like SMVP, gradient or Laplacian.
If GPU and CPU execute essentially the same algorithms, a GPU (Tesla C2070) can outperform a server-class, 6-core
CPU (Intel Xeon X5670) by up to about 4.2 times. We also investigated how the acceleration scales with the problem size
and estimate that the minimum problem size for which GPU can outperform CPU is $\approx 500~000$.
Since our GPU implementation exploits a simple
pressure solver, we compared our results against the CPU running SIMPLE or PISO with a state-of-the art
multigrid (GAMG) pressure solver and found that a better pressure solver is needed for serious CFD applications on GPUs.
We also carried out a detailed, coarse- and fine-grained profiling of our solvers,
finding that their implementation is close to optimal, which confirmed again that the only way
for GPUs to match the efficiency of CPUs in PISO and SIMPLE kernels is a better pressure solver.


\section*{Acknowledgments}
TT and KZ prepared this publication as part of the  project of the City of  Wroc{\l}aw, entitled  ``Green Transfer'' --
aca\-demia-to-business knowledge transfer project co-financed by the European Union under the European Social Fund, under
the Operational Programme Human Capital (OP HC): sub-measure 8.2.1. ZK and MM acknowledge support from
Polish Ministry of Science and Higher Education Grant No.\ N N519 437939.
We kindly acknowledge F.~Rikhtegar and V.~Kurtcuoglu (ETH Zurich) for providing the artery data and fruitful discussions.
This research was supported in part by PL-Grid Infrastructure.
A Tesla C2070 GPU donation from Nvidia is also gratefully
acknowledged.











\end{document}